\documentclass[aps,prl,twocolumn,showpacs,preprintnumbers,amsmath,amssymb,superscriptaddress]{revtex4-1}

\usepackage{graphicx}
\usepackage{dcolumn}
\usepackage{color}
\usepackage{amsmath}
\usepackage[breaklinks,colorlinks,bookmarks=false,citecolor=blue,linkcolor=red,urlcolor=blue]{hyperref}

\begin{document}

\title{Thermal Critical Points and Quantum Critical End Point \\ in the 
Frustrated Bilayer Heisenberg Antiferromagnet}

\author{J. Stapmanns}
\affiliation{Institut f\"ur Theoretische Festk\"orperphysik, JARA-FIT and 
JARA-HPC, RWTH Aachen University, 52056 Aachen, Germany}

\author{P. Corboz}
\affiliation{Institute for Theoretical Physics and Delta Institute for 
Theoretical Physics, University of Amsterdam, Science Park 904, 1098 XH 
Amsterdam, The Netherlands}

\author{F. Mila}
\affiliation{Institute of Physics, Ecole Polytechnique F\'ed\'erale Lausanne 
(EPFL), 
%CH-
1015 Lausanne, Switzerland}

\author{A. Honecker}
\affiliation{Laboratoire de Physique Th\'eorique et Mod\'elisation, CNRS 
UMR 8089, Universit\'e de Cergy-Pontoise, 
%F-
95302 Cergy-Pontoise Cedex, France}

\author{B. Normand}
\affiliation{Neutrons and Muons Research Division, Paul Scherrer Institute, 
%CH-
5232 Villigen-PSI, Switzerland}

\author{S. Wessel}
\affiliation{Institut f\"ur Theoretische Festk\"orperphysik, JARA-FIT and 
JARA-HPC, RWTH Aachen University, 52056 Aachen, Germany}

\date{\today}

\begin{abstract}
We consider the finite-temperature phase diagram of the $S = 1/2$ frustrated 
Heisenberg bilayer. Although this two-dimensional system may show magnetic 
order only at zero temperature, we demonstrate the presence of a line of 
finite-temperature critical points related to the line of first-order 
transitions between the dimer-singlet and -triplet regimes. We show by 
high-precision quantum Monte Carlo simulations, which are sign-free in the 
fully frustrated limit, that this critical point is in the Ising universality 
class. At zero temperature, the continuous transition between the ordered 
bilayer and the dimer-singlet state terminates on the first-order line, giving 
a quantum critical end point, and we use tensor-network calculations to follow 
the first-order discontinuities in its vicinity.  
\end{abstract}

\maketitle

\begin{figure}[t]
\includegraphics[width=0.98\columnwidth]{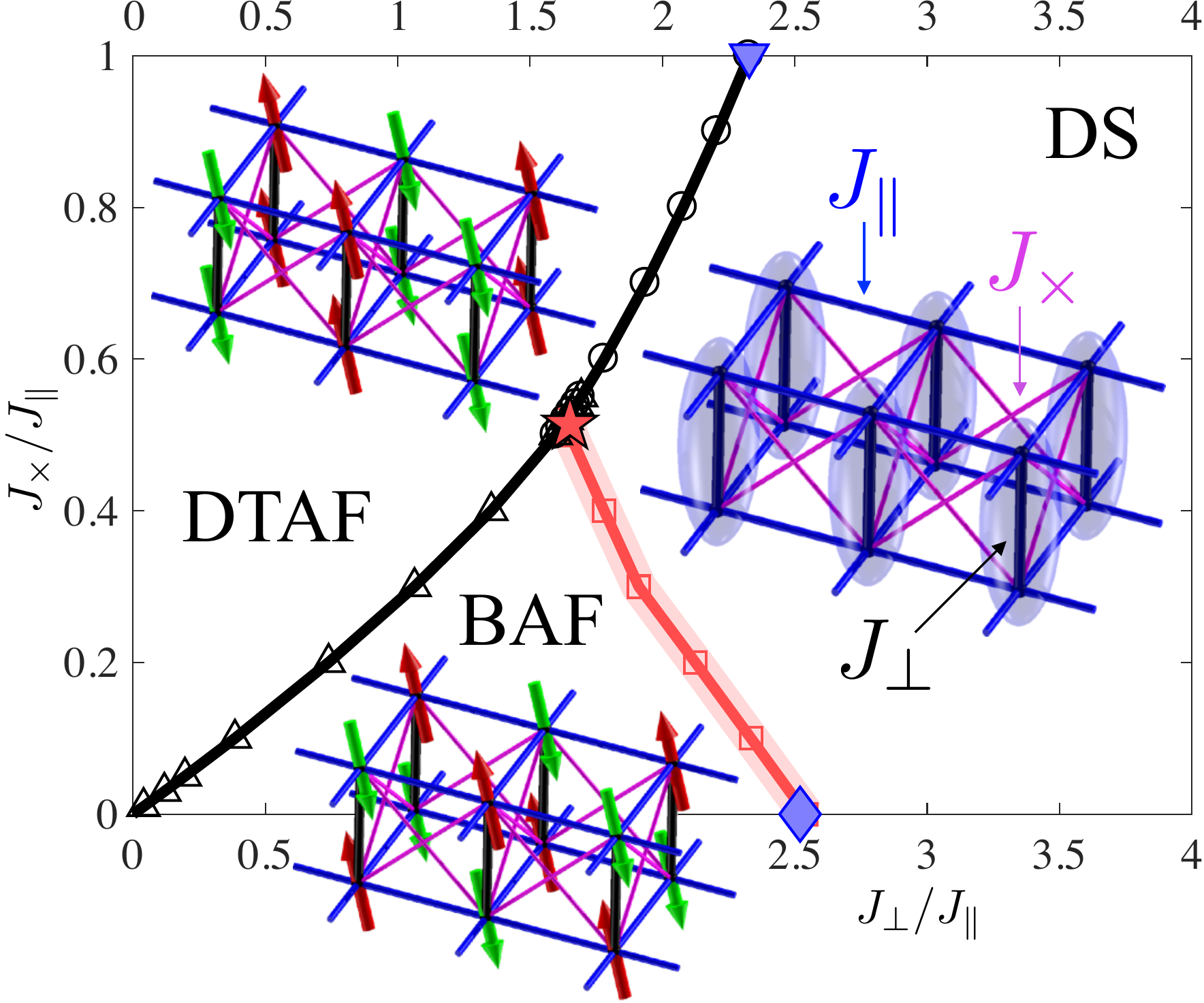}
\caption{Phase diagram of the frustrated Heisenberg bilayer at zero 
temperature. DS: dimer-singlet regime; DTAF: dimer-triplet antiferromagnet; 
BAF: bilayer antiferromagnet. Insets provide schematic representations of 
the three phases, where each site hosts an $S = 1/2$ quantum spin, ellipsoids 
represent singlet states of two spins, and the Heisenberg couplings between 
spins are specified by the parameters $J_\perp$, $J_\|$, and $J_\times$. The 
line of first-order transitions from DTAF to DS or BAF phases is shown in 
black and the line of second-order transitions from the DS phase to the BAF 
in red; red shading indicates the error bars in our calculations. Blue symbols 
denote the QPTs in the unfrustrated (UFB, diamond) and fully frustrated 
(FFB, triangle) bilayers. The red star denotes the QCEP, where the red line 
terminates on the black one.}
\label{fig:s}
\end{figure}

The concept of the critical point is ubiquitous in statistical thermodynamics.
One may need look no further than the liquid-gas transition \cite{rcdlt} 
in systems as familiar as water to find examples where a line of first-order 
transitions terminates as a function of temperature and a control parameter, 
such as the pressure. Because the phase transitions are discontinuous, the 
line has no critical properties, but its termination point does. In contrast 
with this critical point, the term ``critical end point'' (CEP) is reserved 
for the situation where a line of continuous transitions terminates on a line 
of discontinuous ones \cite{rcl,rfu,rfb}. In this case, critical behavior is 
present everywhere on the critical line, and it has been proposed that this 
behavior is reflected in certain properties of the discontinuous boundary on 
which the line terminates \cite{rfb}. 

Quantum spin systems have proven to offer an excellent forum for the 
experimental and theoretical investigation of phase transitions and critical 
phenomena. Quantum phase transitions (QPTs) \cite{rsbook}, predominantly in
low-dimensional systems, have been controlled by pressure \cite{rrnmfmkggmb}, 
applied magnetic field \cite{rgrt,rbtins}, and sample disorder \cite{ryea}, 
and the associated quantum critical regime \cite{rsbook} explored at finite 
temperatures \cite{rmnkbmr}. Frustrated quantum magnets extend the nature of 
the available QPTs to include exact ground states \cite{rmg,rss}, exotic bound 
states \cite{rus1,rus2}, spin liquids \cite{rsb}, and nontrivial topology 
\cite{rrpg}. Here we consider the frustrated bilayer $S = 1/2$ antiferromagnet, 
a two-dimensional (2D) model with Heisenberg exchange. 

In this Letter we demonstrate that, although this system has long-ranged 
magnetic order and spontaneous breaking of SU(2) symmetry only at zero 
temperature, a line of critical points appears at finite temperature, $T$. 
As $T$ is increased, each critical point can be understood as the termination 
of a line of finite-$T$ first-order transitions, exactly like the critical 
point of the liquid-gas transition, and all have Ising nature. The critical 
line is associated with a line of first-order transitions at $T = 0$, where 
we show that the phase diagram as a function of frustration contains a quantum 
critical end point (QCEP), at which a line of continuous transitions 
terminates on the line formed by the first-order quantum phase transitions.

We are motivated by our study of the frustrated spin ladder \cite{rus3}, 
and in particular of its perfectly frustrated limit \cite{rus1,rus2}. Like 
its 1D analog, the frustrated bilayer has a first-order transition between 
dimer-singlet and -triplet regimes, and in the fully frustrated case it has 
completely flat excitation bands composed of many-particle bound states. 
However, in 2D magnetic order is possible at $T = 0$, on top of which thermal 
fluctuations may cause qualitatively different physics to set in. 

The model we investigate is represented schematically in the insets of 
Fig.~\ref{fig:s}. In addition to the interaction, $J_\perp$, defining the 
dimer unit and the intralayer interaction, $J_\|$, defining the two planes 
of the system, we include a symmetrical, diagonal, and frustrating interlayer 
coupling, $J_\times$. Only antiferromagnetic couplings are considered. The 
Hamiltonian for any quantum spin $S$ is 
\begin{equation} 
H \! = \! \sum_{i} \! J_\perp {\vec S}_{i,1} \! \cdot \! {\vec S}_{i,2} + 
\!\!\!\!\!\! \sum_{i, m=1,2 \atop j=i+{\hat x},i+{\hat y} } \!\!\!\!\! [J_\| 
{\vec S}_{i, m} \! \cdot \! {\vec S}_{j, m} \! + \! J_\times {\vec S}_{i, m} 
\! \cdot \! {\vec S}_{j, \overline m}],
\label{eq:essh} 
\end{equation} 
where $i$ is the dimer bond index, $j$ denotes the nearest-neighbor dimers in 
the bilayer, $m = 1$ and $2$ denote the two layers, and ${\overline m}$ is the 
layer opposite to $m$. 

Our initial focus is the fully frustrated bilayer (FFB), $J_\times = J_\|$. In 
this situation, Eq.~(\ref{eq:essh}) can be reexpressed as 
\begin{equation} 
H = J_\| \sum_{i,j} \vec{T}_i \cdot \vec{T}_{j} + J_\perp \sum_{i} 
\bigl[ {\textstyle \frac{1}{2}} \, \vec{T}_i^2 - S\,(S + 1)\bigr] \! ,
\label{eq:exeh} 
\end{equation}
where $\vec{T}_i = \vec{S}_{i,1} + {\vec S}_{i,2}$ is the total spin of dimer $i$ 
\cite{rx,rhmt}. Clearly Eq.~(\ref{eq:exeh}) has one purely local conservation 
law, on $\vec{T}_i^2$, for every dimer in the system. Henceforth we restrict 
our considerations to the case $S = 1/2$. Thus $T_i$ takes the values $0$ [a 
dimer singlet (DS), indicated by the ellipsoids in Fig.~\ref{fig:s}] or 
$1$ [a dimer triplet (DT)]. For a given set $\{T_i\}$, the first term of 
Eq.~(\ref{eq:exeh}) is the Hamiltonian of an open $n$-site spin-$1$ cluster, 
which is nonzero only for groups of $n \ge 2$ neighboring DTs; the second 
term counts these DTs relative to DSs.

As first noted \cite{gelfand91} for the fully frustrated $S = 1/2$ ladder, the 
model of Eq.~(\ref{eq:exeh}) possesses a first-order DS-to-DT QPT as a function of 
the coupling ratio $J_\perp / J_\|$;~the two possible ground states are 
characterized by all $T_i = 0$, when $J_\perp$ is dominant, or all $T_i = 1$ 
when the combination of $J_\|$ and $J_\times$ forces the creation of DTs. For the 
FFB, the ground state in the DT phase exhibits long-range antiferromagnetic 
order of the triplet states, which we denote DTAF. Based on energy arguments 
comparing the DS state with the spin-1 square-lattice Heisenberg model 
equivalent to the DTAF state, this transition is known to occur at $J_{\perp,c}
 = 2.3279(1) J_\|$ \cite{rmhsku}. Several authors have studied this system, 
notably by the construction of exact states \cite{rls,rlss} and in a magnetic 
field \cite{rrdk,rdrhs,rdkr}, and its geometry is realized in the material 
Ba$_2$CoSi$_2$O$_6$Cl$_2$ \cite{rtkoksfumkn,rrkbkd}, albeit with predominantly 
XY interactions. 

\begin{figure}[t]
\includegraphics[width=0.49\columnwidth]{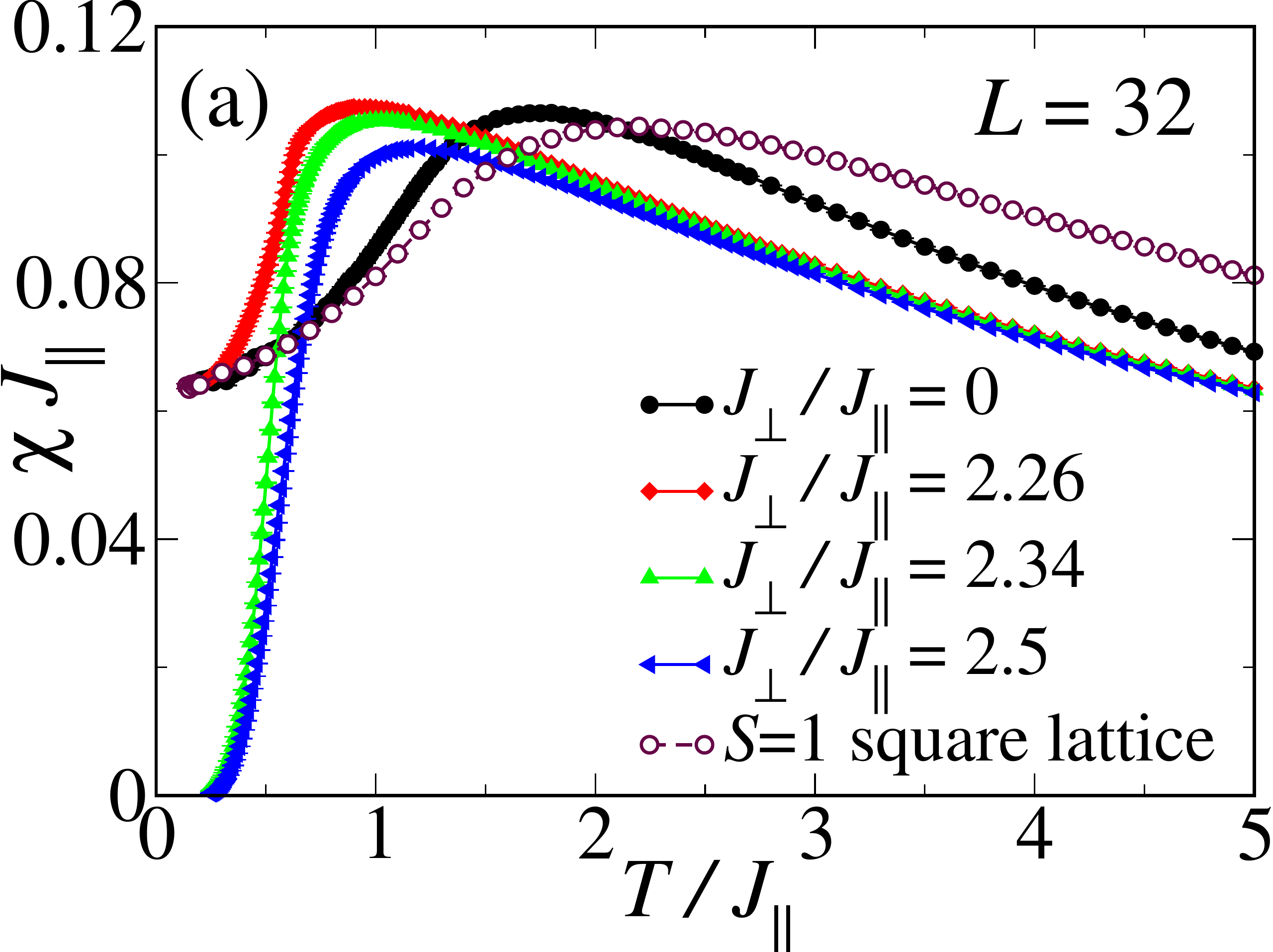}
\includegraphics[width=0.49\columnwidth]{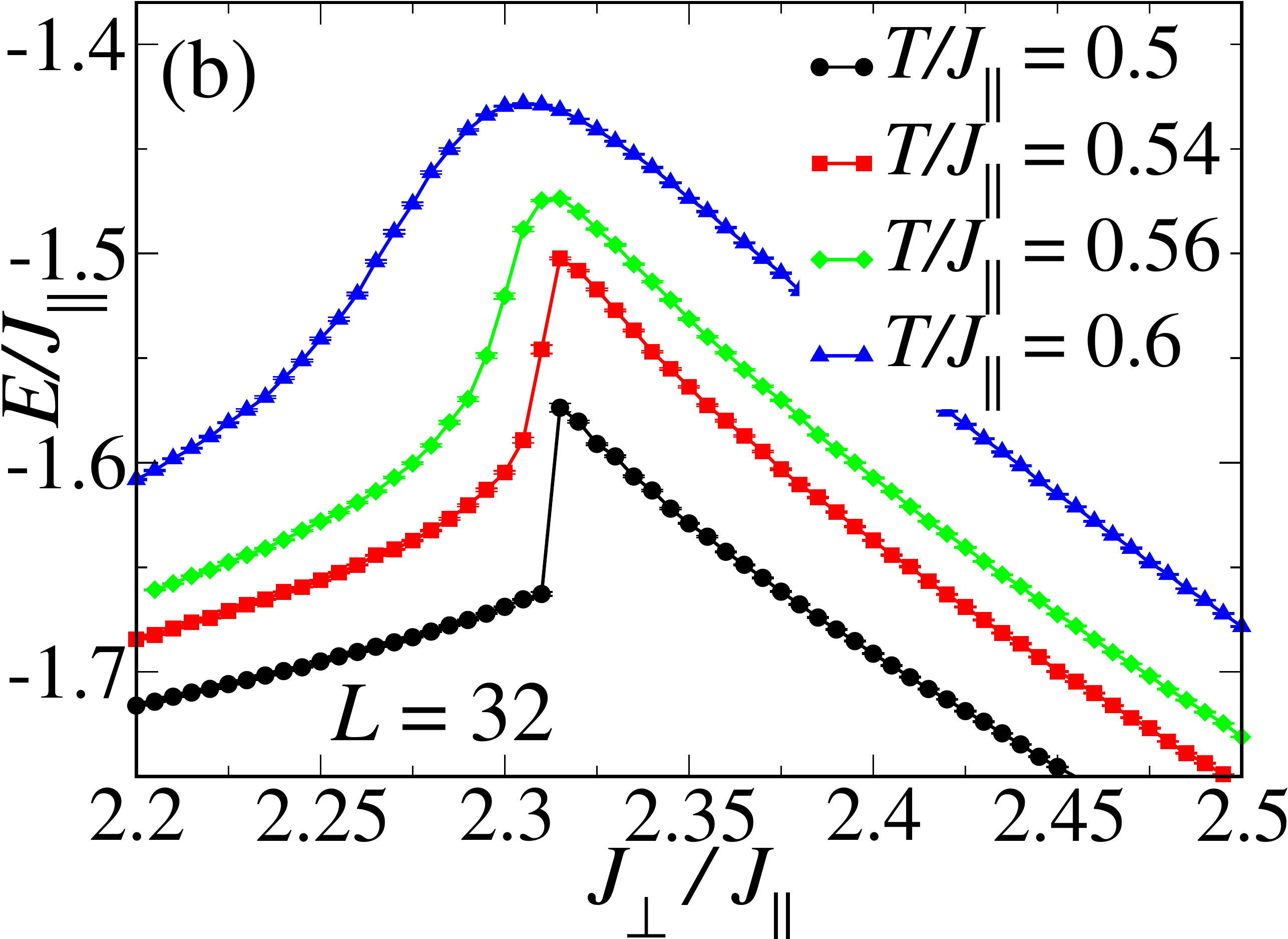}
\includegraphics[width=0.49\columnwidth]{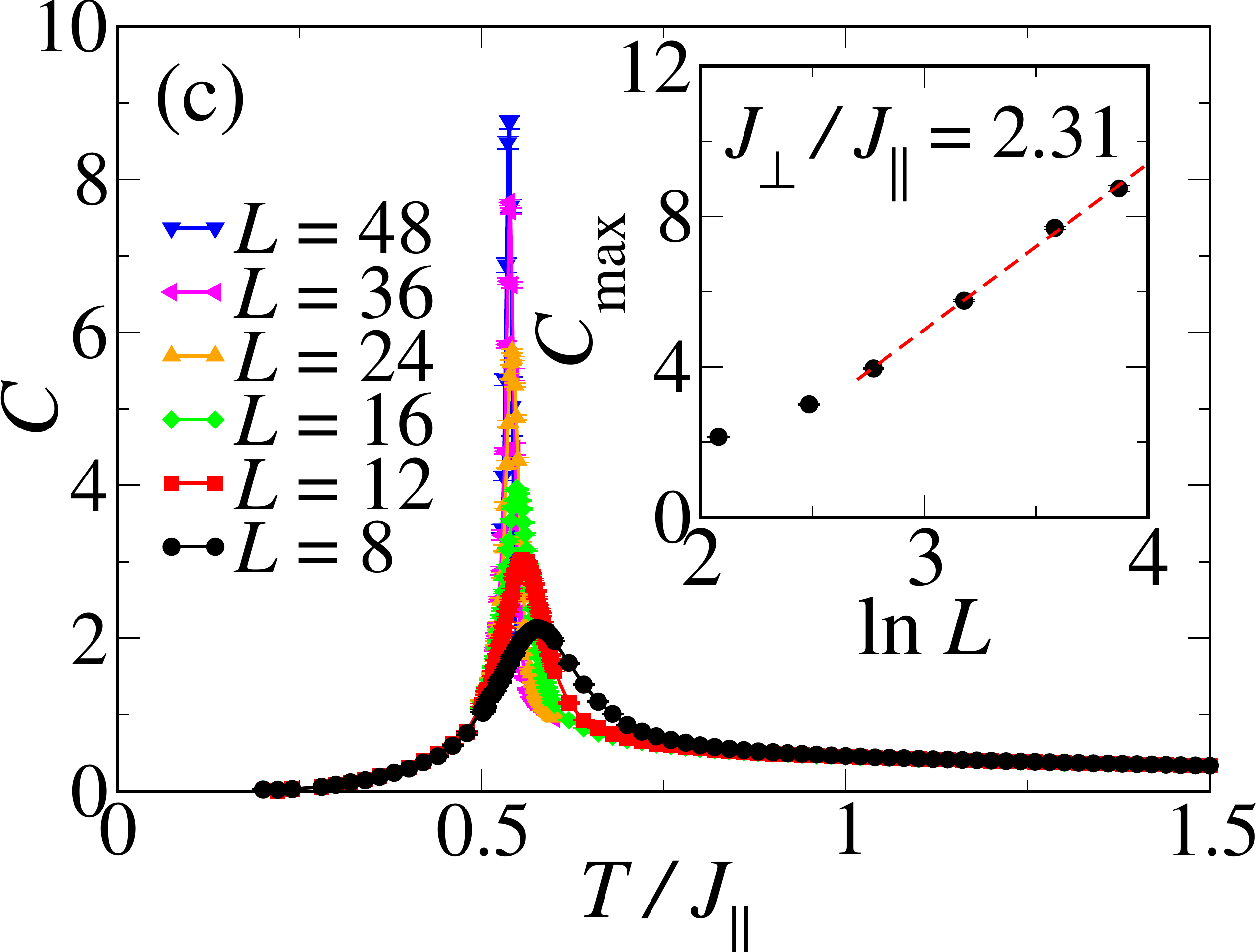}
\includegraphics[width=0.49\columnwidth]{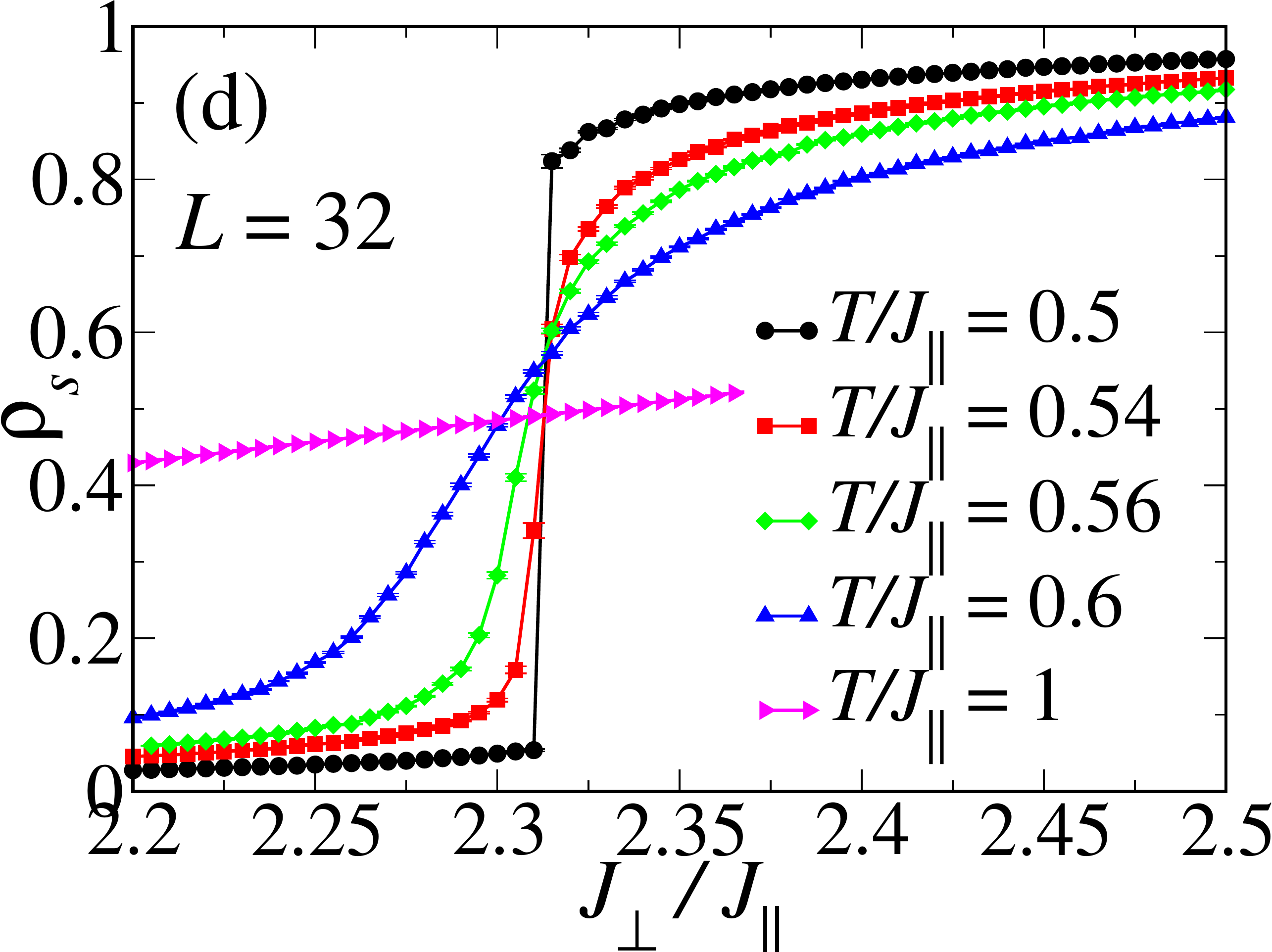}
\caption{Thermodynamic properties of the FFB determined from QMC simulations. 
(a) Magnetic susceptibility, $\chi(T)$, shown for a wide range of coupling 
ratios. (b) Energy, $E$, as a function of $J_\perp/J_\|$ for different 
temperatures. (c) Specific heat, $C(T)$, computed at $J_\perp/J_\| = 2.31$. 
Inset: Finite-size scaling of peak height, $C_\mathrm{max}$. (d) Singlet 
density, $\rho_s$, as a function of $J_\perp/J_\|$ for different temperatures.}
\label{fig:r}
\end{figure}

We use stochastic series expansion \cite{sse0} quantum Monte Carlo (QMC) 
simulations with directed loop updates \cite{sse1,sse2} to examine the 
thermodynamic properties of the FFB in the vicinity of the QPT. It has been 
shown recently \cite{rus1,radp,rus3,rny} that QMC methods can be applied to 
such highly frustrated systems by expressing the Hamiltonian in the dimer basis 
[Eq.~(\ref{eq:exeh})]. The sign problem is entirely absent in perfectly frustrated 
models, including the FFB, and is only moderately serious over a wide range 
of coupling ratios corresponding to imperfect frustration, as we show in 
Sec.~S1 of the Supplemental Material~\cite{SM}. Combined with a parallel 
tempering approach~\cite{rus1}, required to enhance state mixing in the 
vicinity of the first-order QPT, we access system sizes 2$\times$$L$$\times$$L$ 
up to $L = 48$ within the temperature regime relevant for the critical point 
($T \gtrsim 0.3 J_\|$). At lower temperatures, strong hysteresis effects appear 
for couplings close to the QPT. 

The thermodynamic properties obtained from QMC simulations for the FFB are shown in 
Fig.~\ref{fig:r}. The magnetic susceptibility, $\chi(T)$ [Fig.~\ref{fig:r}(a)], 
provides a clear characterization of the gapped DS phase for $J_\perp > 
J_{\perp,c}$, namely an exponentially rapid rise to a broad peak, and of the 
DTAF phase for $J_\perp < J_{\perp,c}$, where $\chi$ approaches a finite value 
at low $T$; this constant is the same as for the spin-1 Heisenberg model on 
the square lattice. The first hint of critical-point behavior is provided by 
the energy [Fig.~\ref{fig:r}(b)], which shows a clear discontinuity as a 
function of the coupling ratio at lower temperatures, but a continuous 
evolution at higher ones. To examine this in more detail we consider the 
dimer singlet density, $\rho_s = \langle N_s \rangle/N_d$, where $N_d$ is 
the number of dimer ($J_\perp$) bonds and $N_s = \sum_i P_{s,i}$ the number 
operator for singlets on these bonds, $P_{s,i}$ being a local singlet 
projector on bond $i$; the DT density is simply $\rho_t = 1 - \rho_s$. 
In the ground state, $\rho_s$ jumps directly from $0$ to $1$ at $J_{\perp,c}$. 
We observe [Fig.~\ref{fig:r}(d)] that this discontinuity persists up to $T 
\simeq 0.54 J_\|$, whereas $\rho_s$ varies smoothly across $J_{\perp,c}$ at 
higher $T$. Thus although magnetic order is found only in the DTAF at $T = 0$, 
the transition from predominantly DT to predominantly DS character persists as 
a first-order line to finite temperatures, of the same order as the interaction 
parameters, before terminating at an apparent critical point.

To rationalize the appearance of critical-point physics, we note that the 
singlet and triplet states on each dimer unit form a binary degree of freedom. 
This effective Ising variable corresponds to the two distinct irreducible 
representations of the spin in the two-site unit cell ($2\otimes 2 = 1 \oplus 
3$). The line of first-order transitions from DS- to DT-dominated states at 
finite temperatures may thus terminate at a finite-$T$ Ising critical point, 
which resembles the liquid-gas transition. This result reflects a key 
additional property of the SU(2)-symmetric frustrated bilayer model. Although 
the continuous symmetry precludes a finite order parameter at $T > 0$, thermal 
fluctuations of the binary variable, whose origin lies in the two-site nature 
of the unit cell, nevertheless stabilize a critical point. 

To identify this Ising critical point in the FFB, we employ finite-size scaling 
of several thermodynamic quantities. In Fig.~\ref{fig:r}(c) we show that the 
specific heat, $C(T)$, computed at $J_\perp \approx J_{\perp,c}$, develops a 
sharp peak at $T \simeq 0.55 J_\|$. The logarithmic form \cite{ro} of the 
size-scaling of the peak height ($C_{\rm max}$, shown in the inset) indicates 
that the transition we observe is consistent with emerging Ising universality. 

\begin{figure}[t]
\centering\includegraphics[width=0.85\columnwidth]{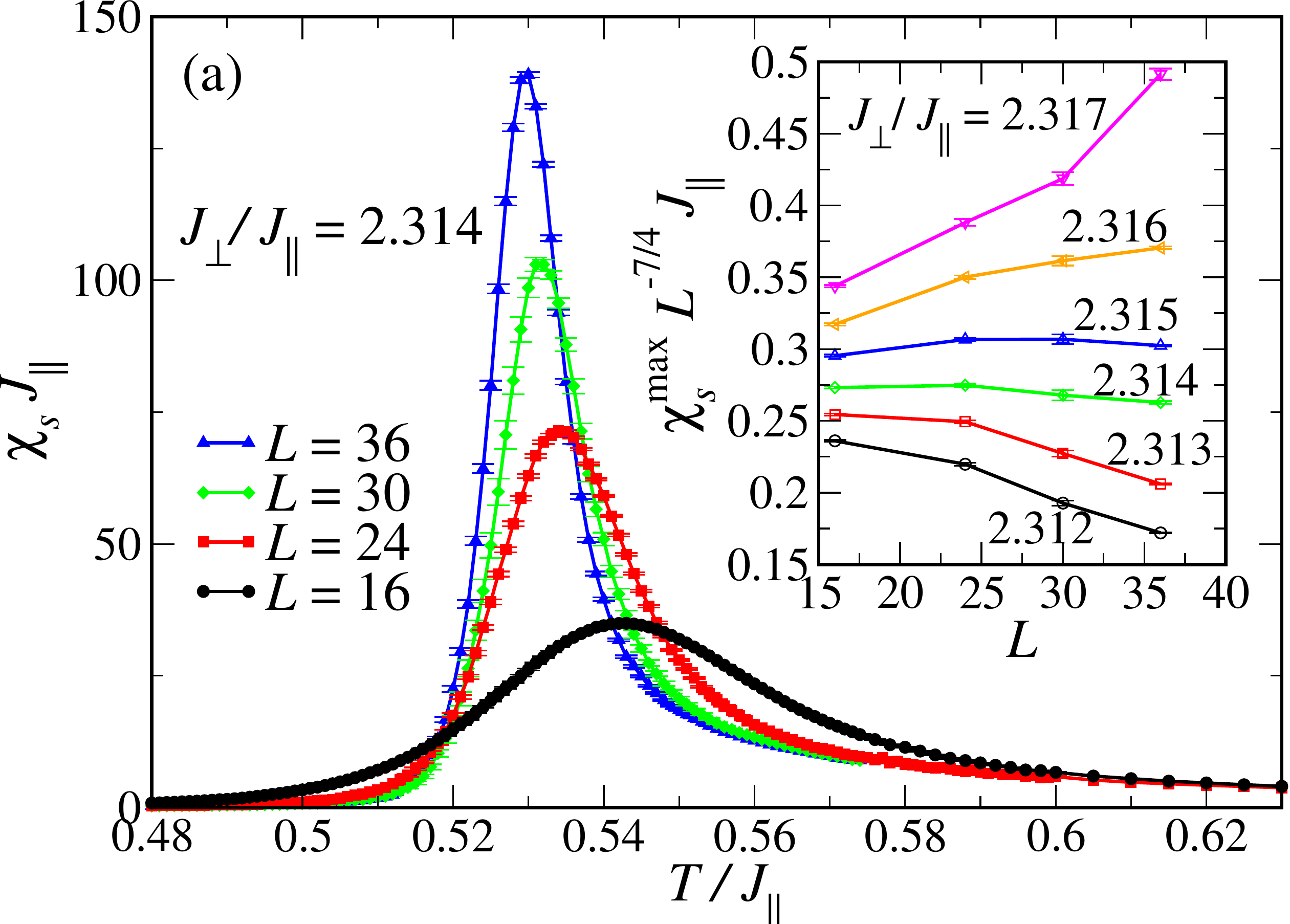}
\centering\includegraphics[width=0.85\columnwidth]{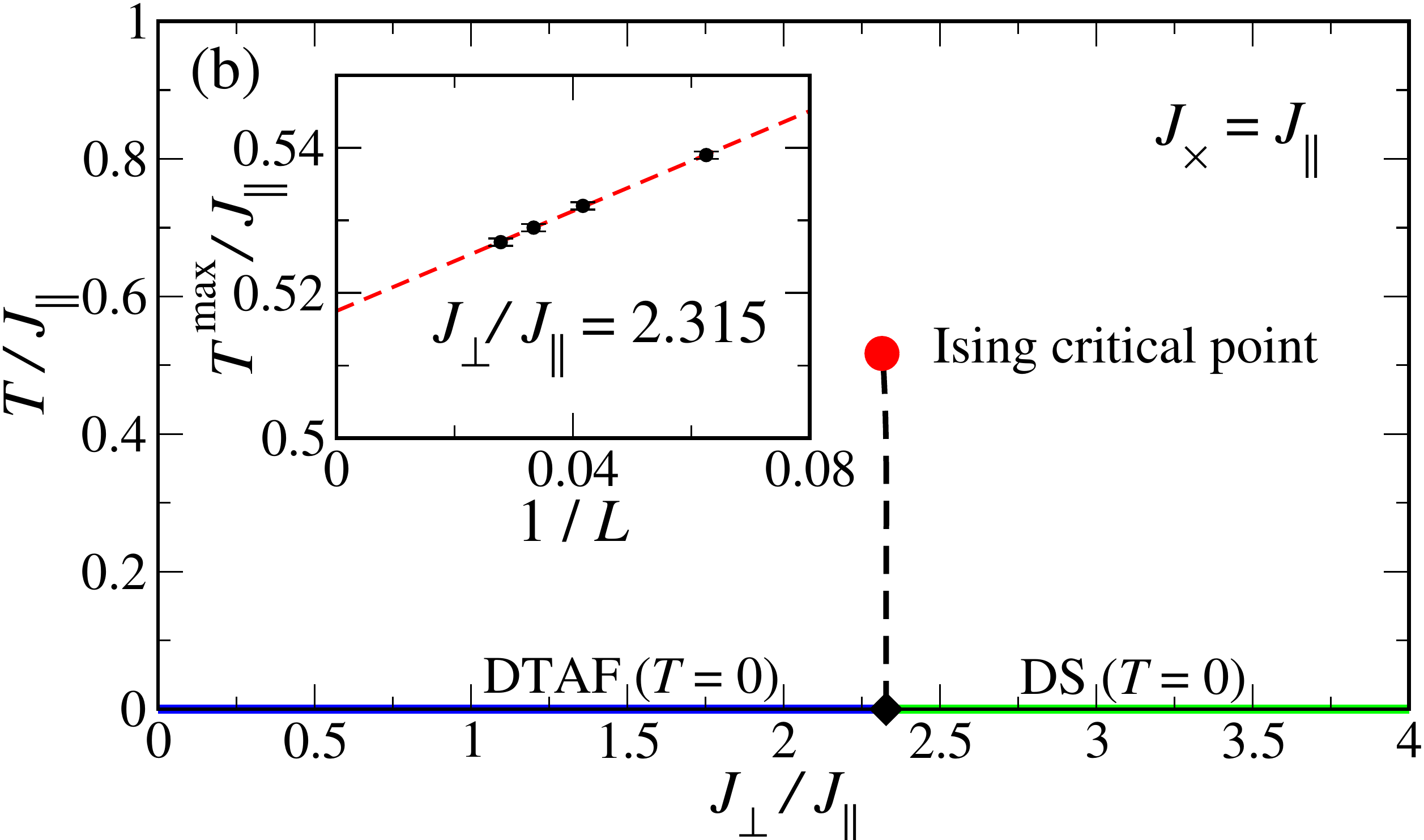}
\caption{(a) Singlet susceptibility, $\chi_s (T)$, computed for systems of 
different sizes, $L$. Inset: finite-size scaling of the rescaled peak height 
for different values of $J_\perp/J_\|$. (b) Phase diagram of the FFB. The dashed 
line marks the finite-temperature first-order transition and the red dot the 
Ising critical point, $(J_{\perp,I},T_I) = (2.315(1) J_\|, 0.517(3) J_\|)$. 
Blue and green colors represent respectively the pure DTAF and DS phases 
at $T = 0$, where the QPT occurs at $J_{\perp,c} = 2.3279(1) J_\|$. Inset: 
Finite-size scaling of the temperature, $T^{\rm max}$, of the peak in 
$\chi_s(T)$ for coupling ratio $J_\perp/J_\| = 2.315$.}
\label{fig:q}
\end{figure}

Our most accurate means of locating the critical point is to compute the 
singlet susceptibility, $\chi_s = \beta/N_d (\langle N_s^2 \rangle - \langle 
N_s \rangle^2)$. Figure \ref{fig:q}(a) shows that $\chi_s(T)$, computed for 
a value of $J_\perp/J_\|$ very near our final estimate of the critical point 
and for a number of system sizes, also shows a sharp peak at the same 
temperature. The inset  shows the dependence on $L$ of the peak maximum, 
$\chi_s^{\rm max}$, scaled by $L^{7/4}$ \cite{rl}, where the curve becoming flat 
(around $J_\perp/J_\| = 2.315$) is in accord with 2D Ising universality. At 
smaller (larger) values of $J_\perp/J_\|$, the rescaled $\chi_s^{\rm max}$ bends 
downwards (upwards) with increasing $L$, indicative of subcritical 
(first-order) behavior.

We draw the coupling-temperature phase diagram of the FFB in 
Fig.~\ref{fig:q}(b). Our estimate of the Ising critical point is 
$(J_{\perp,I},T_I) = (2.315(1) J_\|, 0.517(3) J_\|)$, where $T_I$ is based on 
finite-size scaling of the form $T^{\rm max}(L) - T_I \propto 1/L^{\nu}$, with 
$\nu = 1$ for 2D Ising criticality \cite{rfb67} [inset, Fig.~\ref{fig:q}(b)]. 
Although this first-order line appears to be very steep on the scale of 
Fig.~\ref{fig:q}(b) ($J_{\perp,c} = 2.3279(1) J_\|$ at $T = 0$~\cite{rmhsku}),
its precise shape is a non-trivial consequence of the interplay between 
quantum and thermal fluctuations, which we analyze in Sec.~S2 of the Supplemental Material~\cite{SM}. 

To address the generality of this critical-point phenomenology, we 
consider the bilayer model away from perfect frustration. We first draw the 
ground-state phase diagram connecting the FFB to its unfrustrated counterpart 
(Fig.~\ref{fig:s}). The UFB also has two phases at $T = 0$, an ordered $S = 
1/2$ bilayer AF (BAF) at small $J_\perp$ and a DS phase otherwise. This model 
has been studied extensively, including in Refs.~\cite{rh,rmm,rsasc,rsvb,rwbs}, 
and the QPT is known to occur at $J_\perp/J_\| = 2.5220(2)$ \cite{rwbs}. This 
transition is second-order, with 3D O(3) universality and continuous growth 
of the BAF order parameter, which is quite different from that of the DTAF 
(insets, Fig.~\ref{fig:s}).

\begin{figure}[t]
\includegraphics[width=0.98\columnwidth]{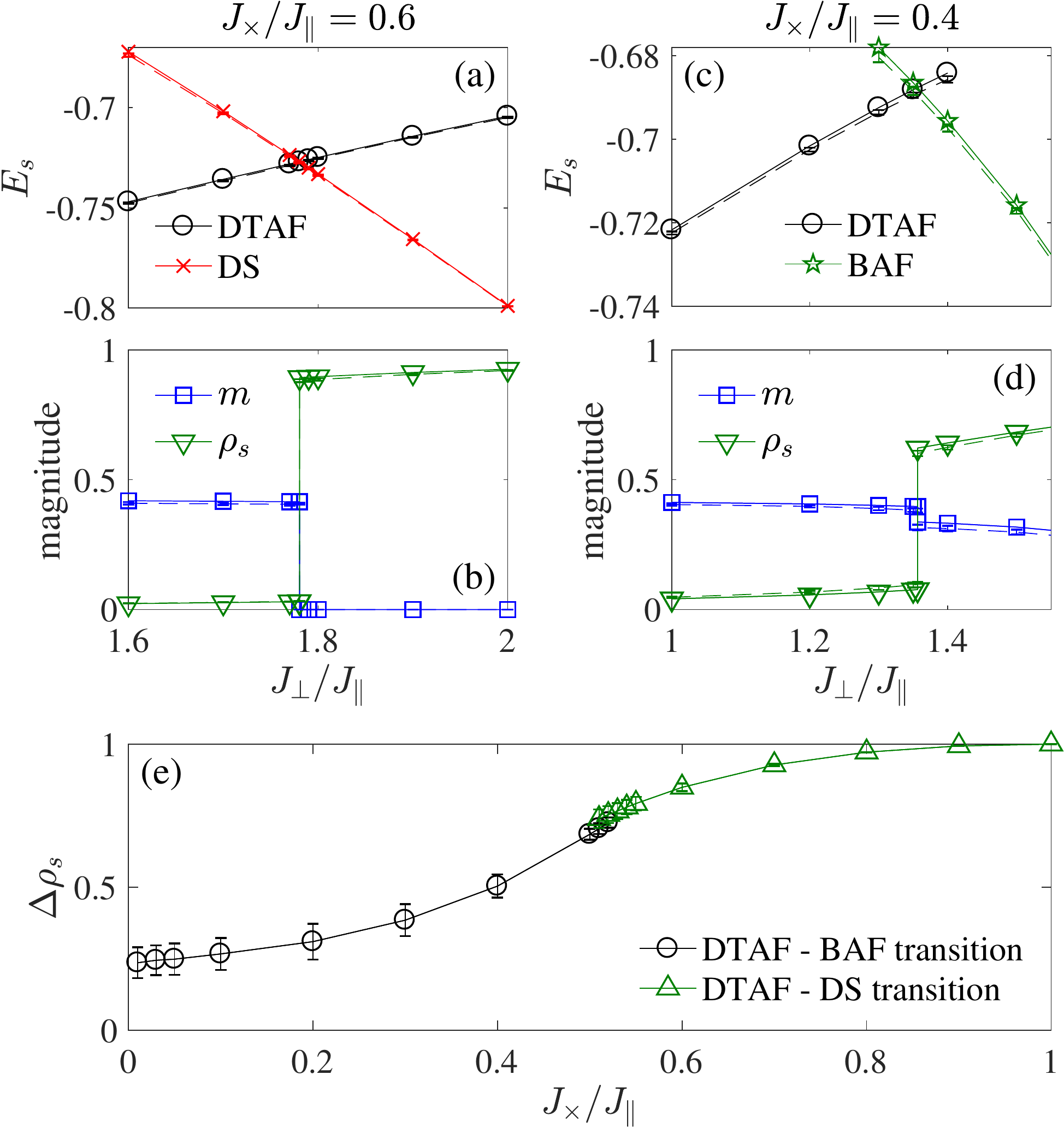}
\caption{(a) Energies of the DTAF and DS phases as functions of $J_\perp / J_\|$ 
at $J_\times / J_\| = 0.6$; the line-crossing marks the first-order transition. 
Full lines show (simple-update) iPEPS results with $D = 10$, dashed lines the 
$D \rightarrow \infty$ extrapolated results. (b) Corresponding singlet density, 
$\rho_s$, and local magnetic moment, $m$. (c)-(d) As for (a)-(b) with $J_\times 
/ J_\| = 0.4$, where the transition is between DTAF and BAF. (e) Discontinuity 
in $\rho_s$, taken from data extrapolated in $D$, shown along the entire 
first-order transition line of Fig.~\ref{fig:s}.}
\label{fig:p}
\end{figure}

We compute the ground-state phase diagram by the method of infinite 
projected entangled pair states (iPEPS)~\cite{verstraete2004,nishio2004,
jordan2008}, which is a variational tensor-network ansatz for a 2D wave 
function in the thermodynamic limit. As we discuss in Sec.~S3 of the 
Supplemental Material~\cite{SM}, the accuracy of this technique can be controlled systematically 
by the bond dimension, $D$, of the tensors, and tensor optimization was 
performed using both the simple- \cite{vidal2003-1,jiang2008} and full-update 
approaches \cite{jordan2008,phien15}. Estimates of energies, singlet densities, 
and magnetic order parameters in the limit of infinite $D$ were obtained by 
extrapolation in $1/D$ \cite{Corboz13_shastry}, as illustrated in the 
Supplemental Material~\cite{SM}. We show our results in Figs.~\ref{fig:p}(a) and \ref{fig:p}(b) for 
a constant frustration ratio $J_\times / J_\| = 0.6$ and in Figs.~\ref{fig:p}(c) 
and \ref{fig:p}(d) for $J_\times / J_\| = 0.4$. A discontinuous transition is 
evident in both cases. 

Critical couplings for the first-order transition line were determined from 
the intersection of the energies of the respective phases [Figs.~\ref{fig:p}(a) 
and \ref{fig:p}(c)]. The second-order transition line was determined from the 
vanishing of the BAF order parameter (obtained by full-update optimization and 
extrapolation). We find that the phase diagram, shown in Fig.~\ref{fig:s}, 
possesses a first-order transition, out of the DTAF phase, for all values of 
$J_\times / J_\|$. The line of continuous BAF-to-DS transitions extends from 
the UFB transition to the point $J_\perp = 1.638(15) J_\|$, $J_\times = 0.520(5) 
J_\|$, where it terminates on the first-order line. By the definition of 
Refs.~\cite{rcl,rfb}, this is a QCEP -- a CEP occurring at $T = 0$. 
The term QCEP has been applied by some authors to field-induced magnetic 
transitions in heavy-fermion systems, apparently to describe critical-point 
physics (termination of a first-order line) \cite{rjunk}, but not in all 
discussions of the same topic \cite{rbkr}.

To our knowledge, there has been very little discussion of the QCEP. In studies 
of the CEP \cite{rfb}, it is proposed that the critical properties of the 
terminating line should be reflected in the properties of the discontinuities 
on the first-order line in the vicinity of the CEP. Unfortunately, we are not 
presently able to perform finite-$T$ calculations in the vicinity of the QCEP 
\cite{SM}. However, from calculations of $\rho_s$, of the type shown in 
Figs.~\ref{fig:p}(b) and \ref{fig:p}(d), we are able to deduce the size of 
the discontinuity, $\Delta \rho_s$, along the first-order line at $T = 0$ 
[Fig.~\ref{fig:p}(e)]. Because $\rho_s$ is related to the energy density, no 
jump is expected in $\Delta \rho_s$ on passing through the QCEP, due to the 
continuous nature of the BAF-DS transition. While $\Delta \rho_s$ is indeed 
continuous within our error bars, our data do suggest a discontinuity in its 
slope across the QCEP. Certainly the critical properties around the QCEP pose 
a challenge to presently available numerical methods.

The limit of weak $J_\perp$ and $J_\times$ is of special interest in the 
frustrated ladder, where the DT-to-DS transition may become continuous 
\cite{rhs} and there have been proposals \cite{roslb} of an intermediate 
phase. In the frustrated bilayer, our calculations show that the first-order 
nature of the transition is robust, with finite jumps in the singlet density 
[Fig.~\ref{fig:p}(e)] all the way to $J_\perp = J_\times = 0$. The value of 
$\Delta \rho_s$ in this limit can be understood from the convergence of 
$\rho_s$ to 1/4 as $J_\perp \rightarrow 0$ in the UFB, where the two layers 
of the BAF become uncorrelated, but the immediate vanishing of $\rho_s$ when 
any finite $J_\times$ at $J_\perp = 0$ stabilizes the triplet state. We conclude 
that the 2D system remains more conventional in this regard than the 1D case.

Returning now to the finite-$T$ Ising critical point, we expect this 
to persist all the way across the phase diagram of Fig.~\ref{fig:s} because 
of its association with the first-order transition. For confirmation, we 
perform QMC simulations at {$J_\times = 0.7J_\|$}, where the sign problem 
remains moderate. As shown in Sec.~S1 of the Supplemental Material~\cite{SM}, our results 
establish that the critical point is still present, occurring at $T = 0.45(1) 
J_\|$. While we are unable by QMC simulations to study the first-order DTAF-to-BAF 
transition line, our iPEPS calculations of $\rho_s$ indicate that the binary 
character of the dimer spin is preserved. We stress that the physics of this 
line of critical points is a consequence not only of the first-order line but 
also of the Ising degree of freedom arising due to the dimer-based unit cell.  

In summary, we have shown that the frustrated $S = 1/2$ bilayer with only 
Heisenberg interactions possesses a line of finite-temperature critical points 
related to a line of first-order transitions in its zero-temperature phase 
diagram. A second line, of continuous transitions from the rung-singlet to 
the bilayer-ordered phase, terminates on the first line, creating a QCEP. 
Understanding the critical properties around the QCEP sets a challenge for 
theory and numerics both in 2D and in higher dimensions. 

\acknowledgments

{\it Acknowledgments.} This work was supported by the Deutsche 
Forschungsgemeinschaft (DFG) in the framework of Grants FOR1807 and RTG 1995,
the Swiss National Science Foundation (SNF), and the European Research 
Council (ERC) under the EU Horizon 2020 research and innovation program 
(Grant No.~677061). We thank the IT Center at RWTH Aachen University and 
the JSC J\"ulich for access to computing time through JARA-HPC.

\end{document}

\title{Supplemental Material for ``Thermal Critical Points and Quantum 
Critical End Point in the Frustrated Bilayer Heisenberg Antiferromagnet''}

\author{J. Stapmanns, P. Corboz, F. Mila, A. Honecker, B. Normand, and  
S. Wessel}

\maketitle

\noindent
{\bf S1. QMC sign problem}
\medskip

The accuracy of our thermodynamic calculations (Figs.~2 and 3 of the main 
text) is attainable because the FFB is amenable to sign-problem-free QMC 
simulations. These are based upon the very simple form of the Hamiltonian 
at $J_\times = J_\|$ [Eq.~(2) of the main text] when expressed in the dimer 
spin basis, 
\begin{equation}
\vec{T}_i = \vec{S}_{i,1} + {\vec S}_{i,2},
\label{eq:defT} 
\end{equation}
of the $J_\perp$ bonds. On moving away from this fully frustrated limit, the 
Hamiltonian takes the form \cite{rus3,radp}
\begin{eqnarray} 
H &=& 
{\textstyle \frac12} (J_\| + J_\times) \sum_{i,j} \vec{T}_i \cdot \vec{T}_{j} +
{\textstyle \frac12} (J_\| - J_\times) \sum_{i,j} \vec{D}_i \cdot \vec{D}_{j}  
\label{eq:defH} \nonumber \\
& & \;\;\;\; + J_\perp \sum_{i} [{\textstyle \frac12} \, \vec{T}_i^2 - S\,(S+1)],
\end{eqnarray}
which for $J_\times \neq J_\|$ contains the spin-difference operators
\begin{equation}
\vec{D}_i = \vec{S}_{i,1} - {\vec S}_{i,2}.
\label{eq:defD} 
\end{equation} 
The $DD$ terms cause the sign problem to reappear. 

As discussed for the frustrated ladder in Ref.~\cite{rus3}, a 1D system with 
only $DD$ (and no $DT$) terms has only a mild sign problem over much of the 
$J_\times \neq J_\|$ phase diagram. This can be traced to the fact that QMC 
configurations with a negative sign are rare with respect to the full 
configuration space and are completely absent for a ladder with open boundary 
conditions~\cite{radp}. For a finite ladder with periodic boundary conditions, 
negative-weight configurations contain a string of bond operators, when 
formulated within the stochastic series expansion, that spans the full spatial 
extent of the system, and the probability for such configurations is suppressed 
strongly by increasing the system size at low temperatures. 

\begin{figure}[t]
\includegraphics[width=1.0\columnwidth]{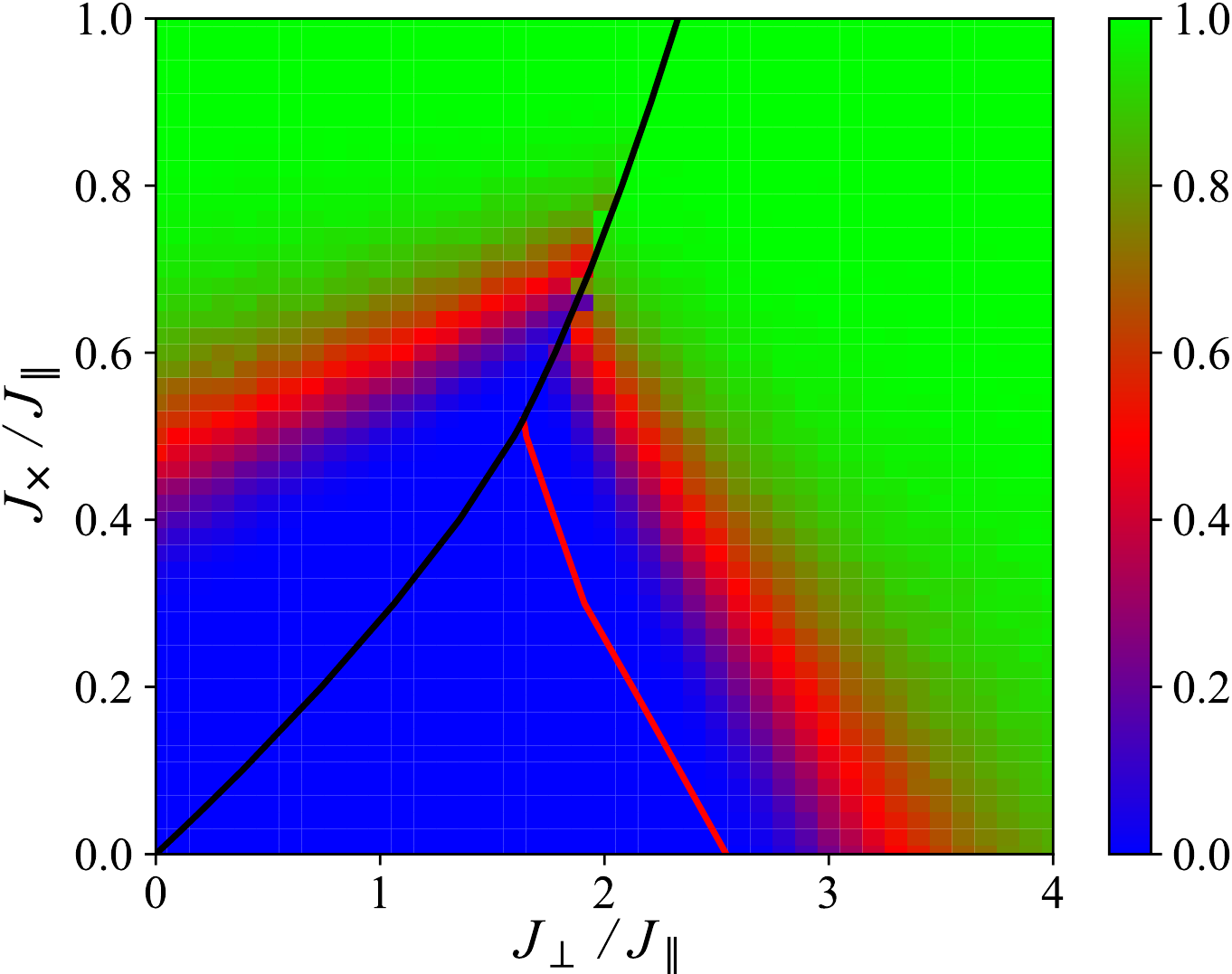}
\caption{\label{spfb} Calculation of $\langle {\rm sign} \rangle$, with 
$L = 10$ and $T = 0.05J_\|$, within a dimer-basis formulation of the QMC 
algorithm. Solid lines reproduce the phase boundaries shown in Fig.~1 of 
the main text.}
\end{figure}

By contrast, negative-weight QMC configurations occurring in the frustrated 
bilayer may contain bond operators from only a small part of the lattice, 
i.e., in 2D these configurations need not span the system and do appear with 
open boundary conditions. Hence we do suffer from strong sign problems in our 
QMC simulations sufficiently far from the fully frustrated limit. To examine 
the onset of the sign problem in more detail, we have performed a scan, over 
the full parameter space of the Hamiltonian in Eq.~(\ref{eq:defH}), to measure 
the average sign, $\langle {\rm sign} \rangle$, of the QMC configurations 
sampled. The results obtained for a moderate system size, $L = 10$, and at 
a temperature of $T = 0.05 J_\|$, are represented by the color scale in 
Fig.~\ref{spfb}. The behavior of $\langle {\rm sign} \rangle$ is correlated 
with the location of the phase boundaries in the ground state (reproduced 
from Fig.~1 of the main text). Although neither the BAF phase nor the 
transitions into it are accessible in the dimer basis, the sign problem 
does remain moderate in large parts of the other two phases.

\begin{figure}[t]
\includegraphics[width=0.95\columnwidth]{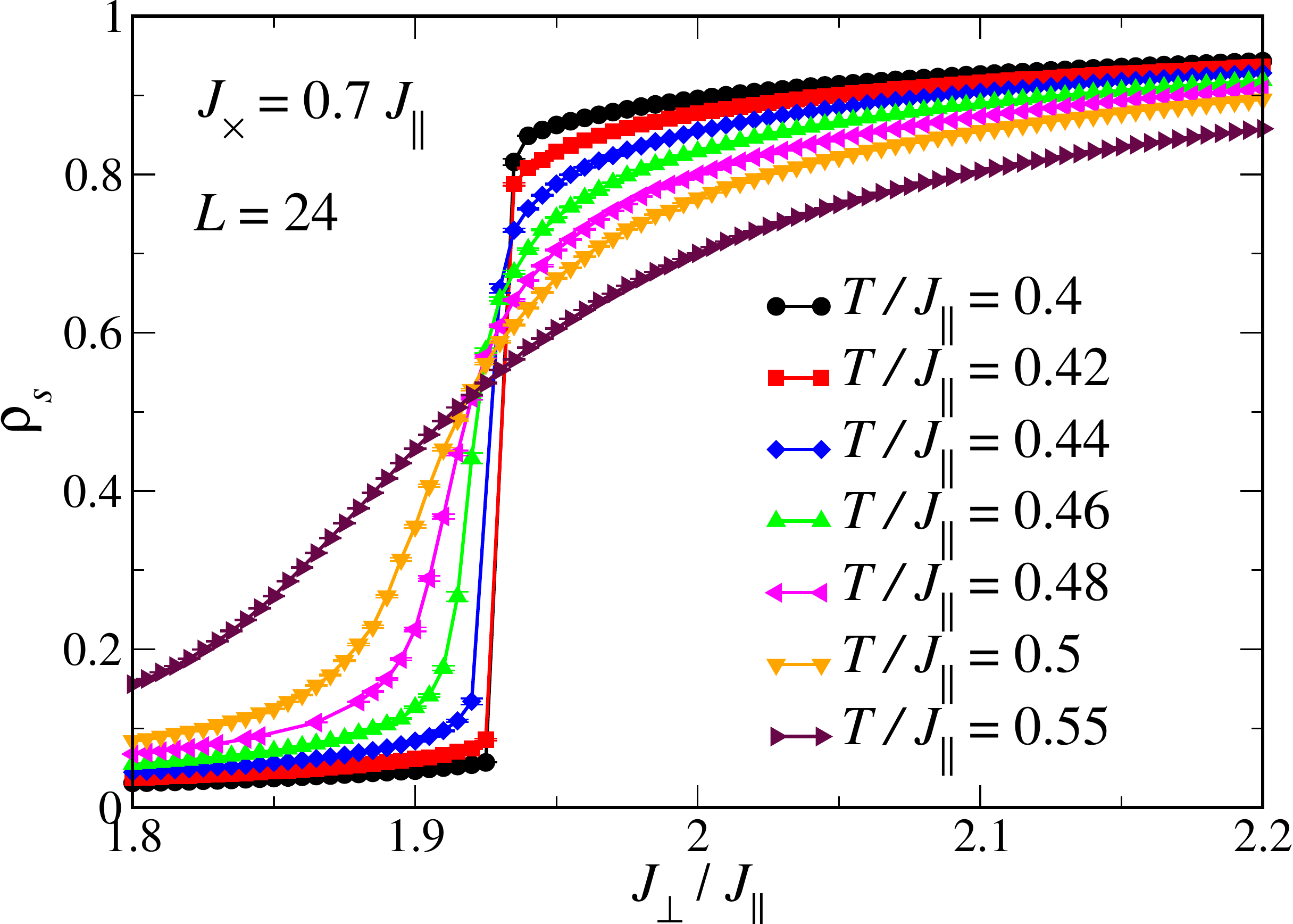}
\caption{\label{jx07} Singlet density, $\rho_s$, calculated on a system of 
size $L = 24$ with frustration ratio $J_\times = 0.7 J_\|$, shown for different 
temperatures as a function of coupling ratio $J_\perp/J_\|$.}
\end{figure}

For practical purposes, a value $\langle {\rm sign} \rangle \gtrsim 0.1$ 
can still be tolerated in our QMC simulations, i.e., compensated by increasing 
the QMC sampling (the CPU time) by a factor of 100. As an example of working 
at partial frustration, we illustrate our estimation of the critical 
temperature at frustration ratio $J_\times = 0.7 J_\|$ (reported in the main 
text). As in Fig.~2(d) of the main text, we computed the singlet density, 
$\rho_s$, at different temperatures as a function of the coupling ratio 
$J_\perp/J_\|$, achieving a maximum system size of $L = 24$. As Fig.~\ref{jx07} 
makes clear, we observe a clear jump in $\rho_s$ near $J_\perp/J_\| = 1.93$ for 
temperatures $T \leq 0.44J_\|$, whereas this quantity exhibits a continuous 
growth for $T \geq 0.46J_\|$. Thus we estimate the location of the critical 
point at $J_\times = 0.7 J_\|$ to be $(J_{\perp,I},T_I) = (1.93(1) J_\|, 0.45(1) 
J_\|)$. We comment that the corresponding first-order transition point in 
the ground state, which we located by our iPEPS calculations, is $J_{\perp,c}
 = 1.9375(8) J_\|$, and hence the finite-$T$ first-order line at $J_\times = 
0.7 J_\|$ shows the same weak $T$-dependence as at full frustration. 

With current computer power it becomes practically impossible to 
perform reliable QMC simulations beyond the regime represented in red in 
Fig.~\ref{spfb}. In the present case, the sign problem prevents us from 
studying the region of the QCEP by QMC, and also from following the line of 
finite-temperature Ising critical points below frustration values $J_\times/J_\| 
\approx 0.7$. For either purpose, accurate thermodynamic studies would require 
system sizes considerably larger than that used in Fig.~\ref{spfb}. For future 
analysis it may be necessary to investigate whether the sign problem can be 
alleviated in the vicinity of the QCEP by further optimizing the local 
computational basis using a more general unitary transformation. We comment 
in closing that the UFB at $J_\times = 0$ is a bipartite system, which can be 
studied in detail by QMC simulations in the conventional site basis without 
encountering a sign problem, and indeed many previous studies of the UFB have 
been performed in this way, including in Refs.~\cite{rsasc,rwbs}. 

\bigskip\smallskip
\noindent
{\bf S2. Finite-temperature transition line} 
\medskip

Here we discuss in detail the form, in the space of coupling ratio and 
temperature, of the finite-$T$ first-order line in the FFB. As noted in the 
main text, in general one expects the level crossing between the DT- and 
DS-dominated phases to move to a different coupling ratio as a function of 
temperature, which influences the two phases quite differently. However, it is 
clear in Fig.~3(b) of the main text that the effects of thermal fluctuations 
are not very strong on the scale of the entire parameter space. 

\begin{figure}[t]
\includegraphics[width=1.0\columnwidth]{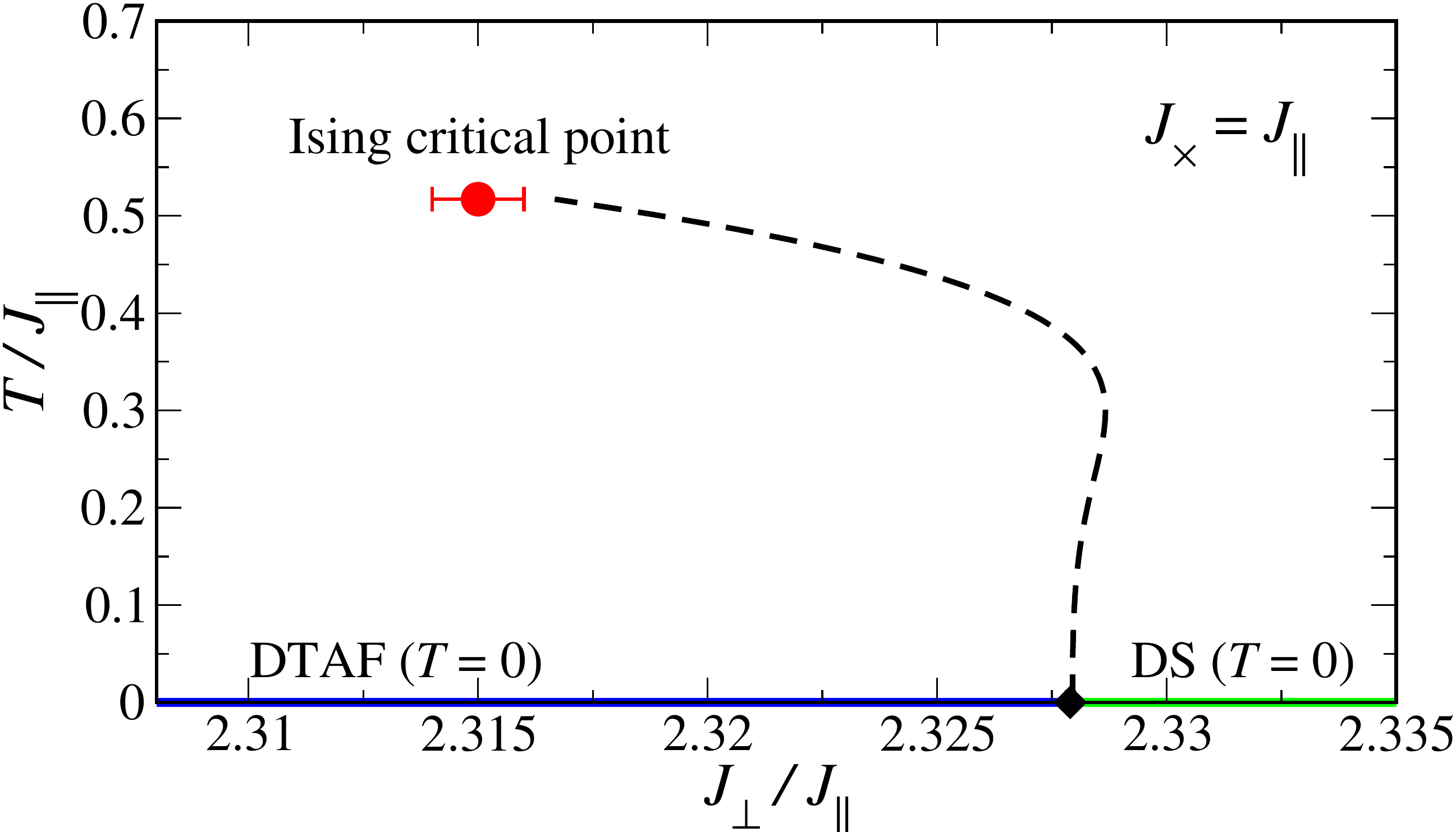}
\caption{\label{ftfloz} Phase diagram of the FFB showing in detail the region 
of the finite-$T$ first-order transition line (dashed). As in Fig.~3(b) of the 
main text, the red dot marks the Ising critical point, $(J_{\perp,I},T_I) = 
(2.315(1) J_\|, 0.517(3) J_\|)$, computed by QMC and the black dot the QPT at 
$(J_{\perp,c},T) = (2.3279(1) J_\|,0)$, taken from Ref.~\cite{rmhsku}. }
\end{figure}

For a full analysis of the problem, we consider the free energies on the two 
sides of the transition line, $F_1(x,T)$ and $F_2(x,T)$, where $x = J_\perp/J_\|$ 
denotes the coupling ratio in the FFB. Let 1 denote the DTAF phase, which is 
gapless, and 2 the DS phase, which is gapped. At finite temperature, $F_i(x,T)
 = E_i(x,T) - T S_i(T)$ ($i = 1$, 2), where $E_1(x,0)$ and $E_2(x,0)$ are the 
ground-state energies and we take $S_i(T)$ to be $x$-independent in the 
vicinity of the QPT. Because the model is in two spatial dimensions, the 
specific heat of the DTAF should obey $C_v(T) = a T^2$, whence by integration 
\begin{equation*}
E_1(x,T) - E_1(x,0) = {\textstyle \frac{1}{3}} a T^3 \;\; {\rm and} \;\; 
S_1(T) = {\textstyle \frac12} a T^2,  
\end{equation*} 
giving to a good approximation
\begin{equation*}
F_1(x,T) = E_1(x,0) - {\textstyle \frac{1}{6}} a T^3. 
\end{equation*} 
In contrast to these power-law expressions, in the gapped DS phase one expects 
only exponentially weak changes in $E_2(x,T)$ and $S_2(T)$. By approximating 
the well-gapped DS state using the response of a single dimer with the same 
spin gap, $\Delta$, we obtain
\begin{equation*}
F_2(x,T) = -  T \ln Z_2(x,T) 
\end{equation*} 
with 
\begin{equation*}
Z_2(x,T) = \exp[-E_2(x,0)/T] + 3 \exp [-(E_2(x,0) + \Delta)/T], 
\end{equation*} 
leading to 
\begin{equation*}
F_2(x,T) = E_2(x,0) -  T \ln [1 + 3 \exp(-\Delta/T)].
\end{equation*} 
By equating the two approximate free energies we obtain the location of the 
first-order transition line from the equation
\begin{equation*}
{\textstyle \frac{1}{6}} a T^3 - T \ln [1 + 3 \exp(-\Delta/T)] = E_1(x,0) - 
E_2(x,0), 
\end{equation*} 
where $a$ is the coefficient of the specific heat per spin-1 entity in the 
DTAF, $\Delta = J_\perp$ is the gap in the DS phase, and $E_1(x,0)$ and 
$E_2(x,0)$ are respectively the zero-temperature energies per dimer of the 
ordered and gapped phases. 

We use the fact that $E_1(x,0) - E_2(x,0)$ is precisely the ground-state 
energy of the $S = 1$ square-lattice antiferromagnet (see Eq.~(2) of
the main text), and take both this 
value and the coefficient $a = 0.2441$ from the QMC simulations of 
Ref.~\cite{rjea}. As Fig.~\ref{ftfloz} makes clear, these considerations 
give an excellent account of the position of the finite-$T$ first-order 
line in the coupling-temperature phase diagram of the FFB. The fundamental 
features of the line are its vertical approach to $J_{\perp,c}$ as $T 
\rightarrow 0$ and the location of the critical point to the left of the 
zero-temperature QPT. However, the latter is actually the result of a 
competition between the power-law terms in $F_1(x,T)$ and the exponential 
behavior in $F_2(x,T)$, which causes the transition line to rise first to 
the right (thermal fluctuations favor DTs) before bending back to the 
left (temperatures higher than $T/J_\| \approx 0.3$ favor the DS phase). 
The crossover temperature appears to depend smoothly on $a$ but to vary 
linearly with $\Delta$, with a coefficient of approximately 0.15.

We stress once again the very narrow regime of coupling ratios over which 
these effects take place in the FFB. Nevertheless, the model provides an 
excellent example of a system in which the interplay of quantum and thermal 
fluctuations can be studied in the context of a key physical consequence, 
namely the location of the line of finite-$T$ transitions. 

\bigskip\smallskip
\noindent
{\bf S3. iPEPS} 
\medskip

Infinite projected entangled pair states (iPEPS) are a variational 
tensor-network ansatz which provides an efficient representation of the 
ground states of 2D systems in the thermodynamic limit~\cite{verstraete2004,
jordan2008,nishio2004}. iPEPS may be considered as a generalization of 
(infinite) matrix-product states to two dimensions. The ansatz consists of 
a unit cell of tensors repeated periodically on the lattice. Here we use one 
tensor per dimer in the bilayer model. Each tensor has one physical index 
representing the local Hilbert space of the dimer and four auxiliary indices 
with bond dimension $D$, which connect to the four nearest-neighbor tensors. 
By using different unit-cell sizes, iPEPS can represent states of differing 
broken translational symmetry. We find that a unit cell with 2 tensors 
arranged in a checkerboard pattern is sufficient to capture all the phases 
of the frustrated bilayer model. 

For the optimization of the iPEPS wave function (i.e., finding the optimal
variational parameters) we use both the simple-update approach~\cite{jiang2008},
in which the truncation of a bond index is based on a local approximation of 
the state, and by the more accurate, but computationally more expensive, 
full-update method \cite{jordan2008,phien15}, where the entire state is taken 
into account for the truncation. Contraction of the tensor network, required 
in the full-update method and to compute expectation values, was performed by a 
variant~\cite{corboz14_tJ} of the corner-transfer-matrix renormalization-group 
method~\cite{nishino1996,orus2009-1}, in which we exploit the U(1) symmetry of 
the model \cite{singh2010,bauer2011}. Further details of iPEPS methods may be 
found in Refs.~\cite{Corboz13_shastry,phien15}.

\begin{figure}[t]
\includegraphics[width=0.98\columnwidth]{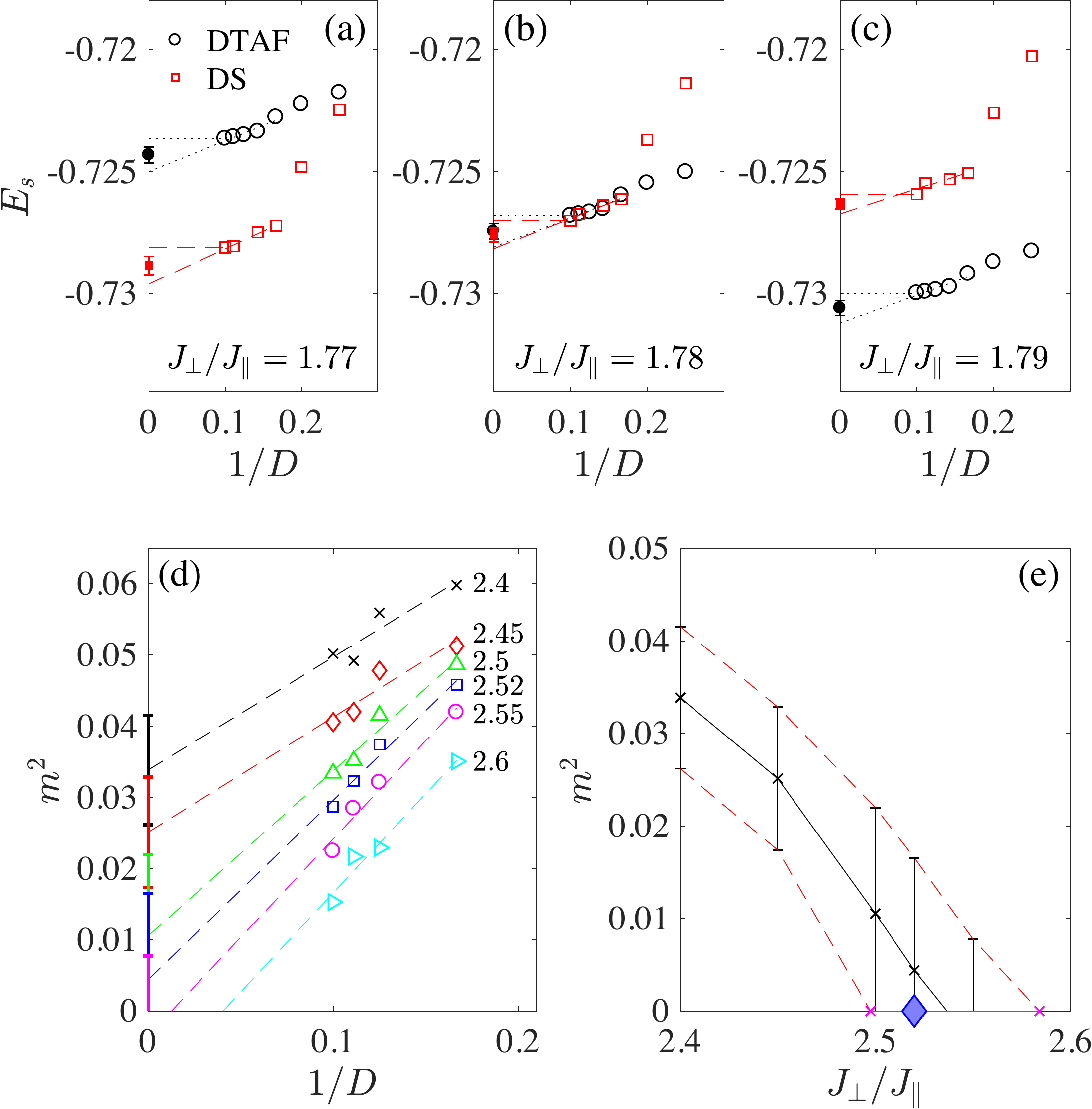}
\caption{\label{ipeps} (a)-(c) Energies of the DTAF and DS states calculated 
by the simple-update method with $J_\times/J_\| = 0.6$ for three different 
values of $J_\perp/J_\|$ and shown as functions of the inverse bond dimension. 
Solid symbols denote the extrapolated estimate in the infinite-$D$ limit. 
(d) Squared local magnetic moment calculated by the full-update method with 
$J_\times = 0$ for the indicated values of $J_\perp/J_\|$ and shown as a 
function of $1/D$. (e) Extrapolated value of $m^2$ obtained from panel (d). 
The blue diamond shows the critical coupling obtained by QMC.}
\end{figure}

To illustrate the procedure by which we obtain estimates of physical 
expectation values from calculations with finite bond dimension, 
Figs.~\ref{ipeps}(a)-\ref{ipeps}(c) show example data for the energies of the 
competing DTAF and DS states close to first-order QPT for $J_\times/J_\| = 
0.6$ [cf.~Fig.~4(a) of the main text]. To estimate the energy in the limit of 
infinite $D$, we take the average between the value obtained by linear 
extrapolation in $1/D$ and the value obtained from our largest $D$ (= 10). 
The former typically provides a lower bound, because the energy converges 
faster than linearly in $1/D$, while the latter is an upper bound. As the 
error bar we take half the difference between the extrapolation and the 
upper bound (an approach similar to Ref.~\cite{Corboz13_shastry}). From the 
intersection of the energy estimates for the two states, taking the error 
bars into account, we deduce a critical value $J_\perp/J_\| = 1.7803 \pm 
0.0014$. For comparison, the value obtained purely from our $D = 10$ data 
is $J_\perp/J_\| = 1.7805$, and thus the finite-$D$ error is very small here. 

In Fig.~\ref{ipeps}(d) we show full-update data for the square of the magnetic 
order parameter in the UFB ($J_\times = 0$). To obtain an estimate in the 
infinite-$D$ limit, here we use a linear extrapolation in $1/D$ and an error 
bar given by half the difference between the extrapolated and largest-$D$ 
values. Although this procedure provides only a rather crude estimate of the 
order parameter, we stress that it does allow us to determine the location of 
the continuous BAF-DS phase transition with reasonable accuracy, as shown in 
Fig.~\ref{ipeps}(e). In this example we find that the order parameter vanishes 
at a critical value $J_\perp/J_\| = 2.54(4)$, which is fully compatible with the 
QMC result, $J_\perp/J_\| = 2.5220(2)$ \cite{rwbs}.